\documentclass[sigconf]{acmart}

\usepackage{tcolorbox}
\usepackage{graphicx}%
\usepackage{listings}
\usepackage{xcolor} 
\usepackage{subcaption}
\usepackage{caption}

\lstdefinelanguage{json}{
    basicstyle=\ttfamily,
    numberstyle=\tiny,
    stepnumber=1,
    numbersep=8pt,
    showstringspaces=false,
    breaklines=true,
    frame=lines,
    backgroundcolor=\color{gray!10},
    literate=
     *{0}{{{\color{blue}0}}}{1}
      {1}{{{\color{blue}1}}}{1}
      {2}{{{\color{blue}2}}}{1}
      {3}{{{\color{blue}3}}}{1}
      {4}{{{\color{blue}4}}}{1}
      {5}{{{\color{blue}5}}}{1}
      {6}{{{\color{blue}6}}}{1}
      {7}{{{\color{blue}7}}}{1}
      {8}{{{\color{blue}8}}}{1}
      {9}{{{\color{blue}9}}}{1},
    keywordstyle=\color{blue},      
    stringstyle=\color{orange},     
    commentstyle=\color{green},     
}

\acmConference[Conference acronym 'XX]{Make sure to enter the correct
  conference title from your rights confirmation emai}
\acmISBN{XXX-X-XXXX-XXXX-X/XX/XX}

\begin{document}


\title{Toward satisfactory public accessibility: A crowdsourcing approach through online reviews to inclusive urban design}



\author{Lingyao Li}
\authornote{Both authors contributed equally to this research.}
\affiliation{%
  \institution{University of South Florida}
  \city{Tampa}
  \state{FL}
  \country{USA}}
\email{lingyaol@usf.edu}

\author{Songhua Hu}
\authornotemark[1]
\affiliation{%
  \institution{Massachusetts Institute of Technology}
  \city{Cambridge}
  \state{MA}
  \country{USA}}
\email{hsonghua@mit.edu}

\author{Yinpei Dai}
\affiliation{%
  \institution{University of Michigan}
  \city{Ann Arbor}
  \state{MI}
  \country{USA}}
\email{daiyp@umich.edu}

\author{Min Deng}
\affiliation{%
  \institution{Texas Tech University}
  \city{Lubbock}
  \state{TX}
  \country{USA}}
\email{mindeng@ttu.edu}

\author{Parisa Momeni}
\affiliation{%
 \institution{University of South Florida}
 \city{Tampa}
 \state{FL}
 \state{Florida}
 \country{USA}}
\email{parisamomeni@usf.edu}

\author{Gabriel Laverghetta}
\affiliation{%
  \institution{University of South Florida}
  \city{Tampa}
  \state{FL}
  \country{USA}}
\email{glaverghetta@usf.edu}

\author{Lizhou Fan}
\affiliation{%
  \institution{University of Michigan}
  \city{Ann Arbor}
  \state{MI}
  \country{USA}}
\email{lizhouf@umich.edu}

\author{Zihui Ma}
\affiliation{%
  \institution{University of Maryland College Park}
  \city{College Park}
  \state{MD}
  \country{USA}}
\email{zma88@umd.edu}

\author{Xi Wang}
\affiliation{%
  \institution{Texas A\&M University}
  \city{College Station}
  \state{TX}
  \country{USA}}
\email{xiwang@tamu.edu}

\author{Siyuan Ma}
\affiliation{%
  \institution{Vanderbilt University}
  \city{Nashville}
  \state{TN}
  \country{USA}}
\email{siyuan.ma@vumc.org}

\author{Jay Ligatti}
\affiliation{%
  \institution{University of South Florida}
  \city{Tampa}
  \state{FL}
  \country{USA}}
\email{ligatti@usf.edu}

\author{Libby Hemphill}
\affiliation{%
  \institution{University of Michigan}
  \city{Ann Arbor}
  \country{USA}}
\email{libbyh@umich.edu}

\renewcommand{\shortauthors}{Li and Hu et al.}

\begin{abstract}
    As urban populations grow, the need for accessible urban design has become urgent. Traditional survey methods for assessing public perceptions of accessibility are often limited in scope. Crowdsourcing via online reviews offers a valuable alternative to understanding public perceptions, and advancements in large language models can facilitate their use. This study uses Google Maps reviews across the United States and fine-tunes Llama 3 model with the Low-Rank Adaptation technique to analyze public sentiment on accessibility. At the POI level, most categories---restaurants, retail, hotels, and healthcare---show negative sentiments. Socio-spatial analysis reveals that areas with higher proportions of white residents and greater socioeconomic status report more positive sentiment, while areas with more elderly, highly-educated residents exhibit more negative sentiment. Interestingly, no clear link is found between the presence of disabilities and public sentiments. Overall, this study highlights the potential of crowdsourcing for identifying accessibility challenges and providing insights for urban planners.
\end{abstract}






\maketitle

\section{Introduction}

As urban populations continue to grow, the importance of creating inclusive cities that cater to all citizens has become increasingly urgent. Despite significant advances in urban planning and development, urban accessibility within cities remains inconsistent \cite{nicoletti2023disadvantaged, spadon2017identifying}. For instance, while public transportation systems are designed to include elevators and audio-visual aids, some regions still have steep stairs and narrow doorways that limit access ~\cite{tatano2017urban}. Similarly, although many newer buildings comply with modern accessibility standards by incorporating ramps, wide doorways, and accessible restrooms, older structures often lack these essential features ~\cite{froehlich2008physical, andani2013investigation}. This inconsistency in accessibility significantly impacts the daily lives of individuals, especially those with disabilities ~\cite{aini2019evaluation, bromley2007city}. Therefore, it is essential to investigate and improve accessibility in urban environments to ensure that cities become truly inclusive for citizens.

Accessibility research has experienced substantial growth in recent decades, with an increasing number of related papers being presented at human-computer interaction (HCI) venues \cite{mack2021we}. The investigation of accessibility within urban environments has often relied on approaches such as surveys ~\cite{lau2004accessibility, ziemke2018accessibility, bekci2023investigation} and field studies ~\cite{wang2023evaluation}. While these approaches offer valuable insights into public perceptions of accessibility and its related service, they are often time-consuming and costly. In addition, they may be limited in scope and participant diversity ~\cite{pinna2021literature}. For example, surveys often struggle to capture a broad range of perspectives and require extra effort to include individuals with disabilities ~\cite{wilson2013accessible}. Field studies, while thorough, can be resource-intensive and may not adequately represent the diverse experiences of users across various areas of a city ~\cite{wang2023evaluation}. As a result, these traditional methods may fail to provide a comprehensive understanding of the accessibility challenges across large urban areas.

Recently, crowdsourcing through social media or online reviews has shown potential for gathering insights from broad and diverse populations ~\cite{neto2018understanding, blohm2013crowdsourcing}. It also highlights the critical role of HCI in today's digital landscape and serves as an interface through which users actively contribute their experiences and perceptions ~\cite{liao2020systematic}. Crowdsourcing from social media or reviews can facilitate data collection and dissemination, making it possible to harness collective intelligence to understand public perceptions of urban environments ~\cite{salomoni2015crowdsourcing, saha2019project}. Therefore, this approach is particularly valuable for gathering wide experiences from citizens. Compared to traditional methods, crowdsourcing through online reviews is also cost-effective, reducing the efforts needed to conduct surveys, on-site inspections, or fieldwork ~\cite{hossain2015crowdsourcing}.

Previous research has highlighted the value of crowdsourced data for investigating traffic flow ~\cite{hu2024multi}, infrastructure management ~\cite{li2022has}, and urban environment ~\cite{li2024crowdsourcing} but rarely focused on inclusive urban design. In addition, while social media data are often unstructured and contain various information, there has been limited exploration of how large language models (LLMs) can be leveraged to understand public perceptions of urban-related issues. Existing studies on using crowdsourcing to evaluate urban accessibility concentrate on specific aspects of the built environment such as walkability ~\cite{gong2023modeling, wang2024investigating} or they investigate small-scale geographic areas, such as individual streets or neighborhoods ~\cite{hara2013combining, saha2019project}. While these studies provide valuable insights, they are limited in scope and do not explain how public perceptions relate to factors such as geography and socioeconomic factors. To build on this literature, we studied two specific research questions:

\begin{itemize}
    \item \textbf{RQ1 (POI analysis).} (1) What patterns exist in public sentiments of accessibility across Point of Interest (POI) types; and (2) What are the key aspects that can explain the sentiments?
    \item \textbf{RQ2 (Socio-spatial analysis).} (1) What patterns exist in public sentiments of accessibility across different geospatial areas; and (2) How are these geospatial patterns associated with local socio-spatial factors?
\end{itemize}

This study explores how user-generated reviews from Google Maps can illuminate public perceptions of accessibility across different POI types and geospatial areas. The first question examines how reviews mentioning accessibility features differ across various POI types and identifies semantic patterns in reviews to discern practical issues that shape public attitudes toward accessibility as positive or negative. The second question explores how accessibility sentiment relates to local socio-spatial factors. The goal is to identify the socioeconomic, demographic, and land development elements that influence public views on accessibility. This research offers valuable insights for urban planners, local policymakers, and accessibility advocates to enhance their decision-making regarding inclusive urban design and to improve accessibility for all, especially individuals with disabilities.


\section{Related Work}

\subsection{Accessibility in urban planning}

Urban environments consist of infrastructure and facilities, such as sidewalks and public transportation, which serve individuals from diverse socioeconomic backgrounds. To ensure equitable access, these environments must cater to people of all abilities~\cite{visualizingUrbanAccessibility}. Urban accessibility can be studied from both macroscopic and microscopic perspectives~\cite{rafael2023urbanAccessibility}. At the macro level, accessibility focuses on large-scale urban planning issues, such as transportation systems and city infrastructure, which affect entire populations~\cite{rafael2023urbanAccessibility}. In contrast, microaccessibility examines the accessibility of individual services and facilities~\cite{grise2019ElevatingAccess}. Prior studies in this area have examined physical barriers in urban infrastructure, such as technical malfunctions in bus entrance/exit ramps~\cite{transportAccessibilityWheelchair} and obstructed sidewalks ~\cite{crowdsourcingGoogleStreetView}. Our study addresses the microaccessibility of urban environments by investigating accessibility challenges at POIs.

As microaccessibility studies frequently highlight, people with disabilities are disproportionately affected by environmental hazards, making them particularly vulnerable to accessibility issues in urban environments~\cite{jampel2018IntersectionsDJ}. For example, a prior study found that when hurricanes struck, households with disabled residents were significantly less likely to evacuate~\cite{brodie2006Katrina}. This underscores the urgency of inclusive urban design. In the United States, the Americans with Disabilities Act (ADA) Accessibility Standards regulate urban infrastructure to ensure that accessibility requirements are met~\cite{ADAAcessibilityStandards}. Additionally, the framework of universal design offers principles for creating environments that accommodate individuals of varying abilities~\cite{designingInclusiveEnvs}. Taught in academic settings~\cite{universalDesign}, these principles have shaped key urban design factors like walkability, legibility, and overall accessibility~\cite{universalDesignUrbanPlanning}.






However, there are still significant gaps in accessibility for disabled people, resulting in decreased quality of life~\cite{NewUrbanAgenda}. For example, poor sidewalk quality has been shown to severely restrict urban mobility for disabled individuals~\cite{mobilityDisabilityUrbanEnv}. Physical barriers often cause anger, anxiety, and frustration for those affected~\cite{transportAccessibilityWheelchair}. Despite the successful passage of ADA legislation, new deployments of public infrastructure may still be inaccessible. For example, newer designs of Bay Area Rapid Transit train cars lacked space for wheelchair users compared to older versions, prompting backlash from disabled communities~\cite{BARTarticle}. Emerging technologies such as micromobility devices and autonomous delivery robots have created additional obstacles on urban pathways~\cite{micromobilityImpact}.

To effectively address these challenges, it is crucial to understand public attitudes toward accessibility. Researchers have employed various methods to assess these attitudes. For example, a typical work conducted interviews with 33 disabled participants to investigate their perceptions of public facilities~\cite{fange2002accessibility}. Surveys are also popular tools to assess people's attitudes toward urban vitality~\cite{urbanVitalityIstanbul} and accessible transportation~\cite{accessibleTransportTurin}. Through surveys, prior work found that people, politics, and budget were intrinsically tied to underfunded accessibility improvement projects ~\cite{saha2021urban}. However, these traditional methods, based on either surveys or interviews, are hindered by their limited geographic scope and lack of scalability. New approaches using crowdsourcing may offer an additional solution to these issues of limited data~\cite{ilieva2018SMD}.

\subsection{Crowdsourcing to investigate accessibility}

\subsubsection{Crowdsourcing to study urban environments}
Crowdsourcing approaches have become increasingly prevalent in studying urban environments, providing researchers with innovative tools for real-time, large-scale, decentralized data collection and analysis~\cite{certoma2015crowdsourcing}. This trend reflects a growing recognition of the potential for crowdsourcing to address complex urban challenges ~\cite{velazquez2024crowdsourcing}. In the context of urban accessibility, crowdsourcing has demonstrated its value. For instance, previous research has leveraged crowdsourcing to improve accessibility for mobility-impaired individuals in smart cities, illustrating how these efforts are driving the development of more inclusive urban spaces~\cite{panta2019improving}. However, there is a need to differentiate between two primary types of crowdsourcing approaches: (1) active crowdsourcing, and (2) passive crowdsourcing, both of which offer unique contributions to urban planning, accessibility, and sustainability~\cite{kamaldin2019smartbfa}.

\subsubsection{Active crowdsourcing}
Active crowdsourcing relies on direct participation from individuals, where users consciously contribute data or feedback~\cite{mobasheri2017wheelmap}. For example, public webcams and citizen science initiatives have played a pivotal role in enhancing flood models and early warning systems through real-time data collection~\cite{helmrich2021opportunities}. In the context of urban accessibility, active crowdsourcing is also effective. A typical example is Project Sidewalk, which allows users to annotate street-level images from Google Street View (GSV) to identify accessibility barriers. Using a gamified interface on Amazon Mechanical Turk, Project Sidewalk efficiently scales up accessibility audits, demonstrating the advantages of active crowdsourcing for urban accessibility audits~\cite{crowdsourcingGoogleStreetView, saha2019project}. Another typical application is The EasyGo platform, which exemplifies active crowdsourcing by involving users to report and map barriers and facilities across cities. This platform offers custom routing for mobility-impaired individuals \cite{panta2019improving}. However, these active crowdsourcing approaches often face challenges related to data sparsity and user fatigue~\cite{ding2014survey}.




\subsubsection{Passive crowdsourcing}

In contrast, passive crowdsourcing does not require direct user involvement \cite{loukis2015active}. Instead, it leverages data generated from everyday activities and passive systems. Compared to active crowdsourcing, passive crowdsourcing has several typical advantages. First, passive crowdsourcing does not require direct user involvement and allows for continuous data collection without the need for participant recruitment, which can be resource-intensive and limiting in scale ~\cite{ghermandi2019passive, ding2014survey}. More importantly, passive crowdsourcing can help gather vast amounts of data from a broad set of populations, providing richer and more representative datasets. This broader data collection also enables deeper and more nuanced analysis, revealing insights that might be missed with smaller, actively gathered datasets from conventional methods like surveys or interviews in urban studies ~\cite{li2024crowdsourced, chuang2022effects}.




Previous research has shown the remarkable potential of passive crowdsourcing via GPS tracking or social media to decipher public sentiments and attitudes across domains in urban environments ~\cite{basak2020crowdsourcing}. For instance, Williams et al. (2019) ~\cite{williams2019ghost} gathered data from several Chinese social media platforms, such as Baidu Map, and created a computational model using residential visit patterns to identify ``Ghost cities'' in China—areas with significant housing vacancies due to excessive development relative to their population size. Chen et al. (2019) ~\cite{chen2019identifying} analyzed Facebook data to study urban vibrancy in Hong Kong, focusing on residents' visits to POIs to understand spatial structures and leverage social media insights for enriched urban planning. Similarly, Chuang et al. (2022) ~\cite{chuang2022effects} collected over 20 million geolocated tweets from July 2012 to October 2016 to analyze park visits in Singapore. Their statistical analysis showed that high surrounding population density and family-oriented facilities, such as playgrounds, could lead to greater visitor density in parks. 

Passive crowdsourcing has also illustrated its potential in addressing urban environmental issues or emergent events, including disaster damage estimation ~\cite{li2023exploring}, community resilience ~\cite{ma2024investigating}, and evacuation efforts ~\cite{li2021data}. For example, air pollution monitoring has been transformed by mobile sensors on vehicles and smartphones, enabling the collection of detailed air quality data at a city-wide scale~\cite{huang2018crowdsource}. Another typical study utilized the passively collected location data from mobile devices to examine the impact of floods, winter storms, and fog on road traffic ~\cite{hu2024multi}. It found that the patterns that emerged from crowdsourced data could provide meaningful insights into disaster response and recovery in road transportation systems. 

Existing efforts to use passive crowdsourcing for urban accessibility have been limited. One typical application is the SmartBFA system ~\cite{kamaldin2019smartbfa}, which gathers data through Internet of Things (IoT) devices attached to wheelchairs. Another key initiative is AccessMap ~\cite{aly2021better}, which enhances road maps by passively crowdsourcing accessibility data from smartphone sensors to detect features like ramps and curb cuts. Additionally, previous research on automated road accessibility assessments like mPass has utilized wheelchair-mounted sensors for passive crowdsourcing ~\cite{iwasawa2015toward, prandi2014mpass}. These sensors collect accelerometer data to evaluate ground surface conditions and identify obstacles such as steps and slopes. However, these efforts often remain limited to small-scale experiments or require the installation of specialized sensor equipment.

Prior studies have demonstrated the significant potential of crowdsourcing to enhance urban studies. However, several challenges remain. First, many previous studies focus primarily on the number of social media visits to infer urban environment or accessibility, often overlooking the rich textual content shared by users that could reveal deeper insights into their sentiments and opinions. Second, there is a noticeable gap in research specifically addressing public perceptions of accessibility for disabled individuals. This area is complex and requires nuanced understanding to decode opinions that go beyond simple visitation metrics. Third, given that social media posts and online reviews are unstructured, there has been limited exploration into using LLMs to improve the interpretation of these opinions. To address these challenges and harness the potential of passive crowdsourcing, we propose the two research questions outlined in the introduction.

\section{Data and Methods}

\begin{figure*}[htbp]
  \centering
  \includegraphics[width=0.98\textwidth]{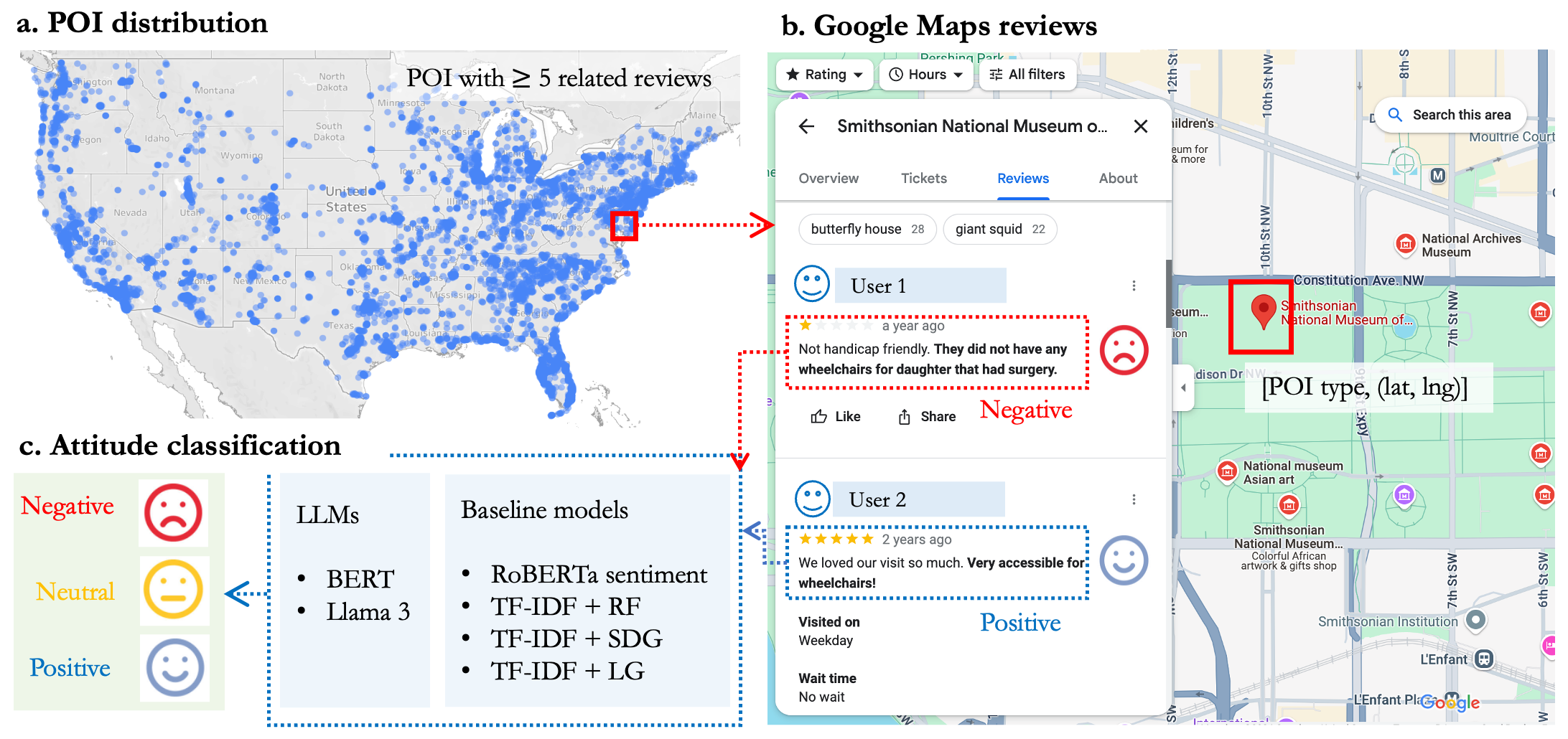}
  \caption{An illustrative framework to conduct the data analysis.}
  \label{fig:framework}
\end{figure*}

Figure \ref{fig:framework} illustrates a framework for conducting research collected from Google Maps reviews, specifically focusing on accessibility-related sentiment analysis. Figure \ref{fig:framework}a shows the distribution of POIs with 5 or more related reviews in the United States, which are then used for subsequent POI analysis. Figure \ref{fig:framework}b displays the Google Maps reviews for the Smithsonian National Museum. It includes two example user reviews ---one negative and one positive--- related to accessibility for wheelchair users. After we collect those reviews, we build models for classifying the opinions from Google Map reviews (Figure \ref{fig:framework}c). The classification process involves using both LLMs like BERT and Llama 3, as well as baseline models such as RoBERTa sentiment analysis and various TF-IDF-based models. 

\subsection{Data preparation}

Google Maps reviews serve as our primary source for gauging residents' attitudes toward accessible facilities. Google Maps allows users to freely rate and review POIs, including businesses, attractions, and public spaces. We select Google Maps reviews as our primary data resource due to two major reasons (see example in Figure \ref{fig:framework}b). First, it has seen substantial growth in user reviews, outperforming competitors like Yelp and TripAdvisor \cite{munawir2019visitor}. Second, unlike social media platforms such as Facebook or Twitter (now called X), Google Maps reviews specifically focus on customer experiences with businesses and locations, making them particularly suitable for our crowdsourcing approach. The online reviews collected from Google Maps can provide a reliable and relevant dataset for our study on attitudes toward accessible facilities \cite{lee2018assessment}.

To implement the study, we utilize the dataset published by UC San Diego \cite{li2022uctopic, yan2023personalized}, which encompasses Google Maps reviews up to September 2021 across the United States. The dataset consists of 666,324,103 reviews with 4,963,111 POIs covered. For each POI, this data repository includes:

\begin{itemize}
    \item User-generated review data: Usernames, ratings, comments, and images.
    \item Business metadata: Addresses, geolocation data, descriptions, categories, pricing, operating hours, links to the business, and other related information.
\end{itemize}

To filter reviews that suggest residents' attitudes toward accessible facilities, we refer to the ADA Accessibility Guidelines (ADAAG) \cite{ADAAG, ADAAcessibilityStandards}. These guidelines list accessible elements and spaces, along with their respective requirements. After reviewing the guidelines and filtering online reviews using each keyword, we finalize the following search list, as presented below.

\begin{tcolorbox}[colback=gray!10!white, colframe=gray!50!gray, halign=left, top=0mm, bottom=0mm, left=1mm, right=1mm, boxrule=0.5pt]
\fontsize{9pt}{9pt}\selectfont 
\linespread{1}\selectfont
\textit{Search list: ``accessible,'' ``accessibility,'' ``ada compliance,'' ``ada compliant,'' ``blind,'' ``braille,'' ``curb cut,'' ``curb ramp,'' ``deaf,'' ``disabilities,'' ``disability,'' ``disabled,'' ``grab rail,'' ``hand splint,'' ``handicap,'' ``handrail,'' ``hearing loss,'' ``induction loop,'' ``mobility aid,'' ``service animal,'' ``service cat,'' ``service dog,'' ``tactile map,'' ``tactile paving,'' ``tactile warning,'' ``tgsi,'' ``vision impairment,'' ``visual impairment,'' and ``wheelchair.''}
\end{tcolorbox}

Several points need to be highlighted given this search list. First, it should be noted that some elements mentioned in the ADA Accessibility Guidelines, such as ``elevator,'' ``open area,'' and ``play area,'' do not return any relevant Google Maps reviews in this context. Including such elements could introduce a significant amount of noise into our analysis, and therefore, we exclude them from our search list. As a result, we filter in 1,013,771 reviews that potentially reflect attitudes toward accessible facilities. While it's important to note that online reviews may contain typographical errors, this study does not account for these typos due to the challenges in comprehensively addressing all possible misspellings for each term. Last, it is worthwhile noting that our search strategy captures different forms of words. For instance, both ``handicap'' and ``handicapped'' are included in our results, as our filtering process identifies reviews containing the root word, thus encompassing various grammatical forms. 


\subsection{Data annotation}

Identifying users' perspectives on accessibility from Google Maps reviews can be approached as a text classification problem in natural language processing (NLP). Initially, we explore the possibility of using pre-existing sentiment analysis tools, such as RoBERTa sentiment model. However, phrases such as ``handicap accessible'' or ``Not ADA accessible'' are frequently categorized as neutral by these models. This likely occurs since general sentiment analysis tools fail to capture the subtle nuances related to attitudes toward accessible facilities. As a result, it becomes necessary to create specialized text classification models tailored to this specific research context.

Developing classification models requires training and testing datasets. Typically, attitude classification encompasses three categories: positive, neutral, and negative. However, we observe that certain reviews, despite containing pertinent keywords, do not express any specific attitudes toward accessible facilities or services. For instance, statements like ``The restaurant is in a good location and very accessible to the freeway'' or ``It is well located and accessible'' are not related to accessible facilities in the intended context. This is because keywords such as ``accessible'' can have different meanings depending on the context. To address this, we introduce an additional class, ``unrelated,'' to the existing categories. Generally, comments that mention the availability or satisfaction of accessible facilities are categorized as positive. Comments that highlight the lack of accessible facilities, poor quality of such facilities, or inadequate service are classified as negative.

Specific examples of each category are provided in the Table \ref{tab:example} below. We chunk the reviews into sentences and then focus on the part that contains the targeted aspects for attitude classification. This is because some online reviews are long and may cover aspects unrelated to attitudes toward accessible facilities or services. This handling can be useful when a review expresses conflicting sentiments about various aspects (see the fourth example in Table \ref{tab:example}).

\begin{table*}[htbp]
  \caption{. Representative Google Maps reviews for public attitudes toward accessibility}
  \centering
  \begin{tabular}{p{0.55\linewidth}p{0.25\linewidth}p{0.1\linewidth}}
    \hline
    \textbf{Google Maps review [sic]} & \textbf{Targeted text [sic]} & \textbf{Attitude} \\
    \hline
    UNBELIEVABLE! So many cool Great Danes! And what a valuable service for people with disabilities! Thank you for helping others! & And what a valuable service for people with disabilities! & Positive \\
    \hline
    Such a great clean park. Not much for kids to do but it has a very nice walking trail around the park. Good level parking lot with handicap space. Would probably be a great place to walk your dog or just hang out with loved ones. Benches available for sitting but no covered areas if its raining. & Good level parking lot with handicap space. & Positive \\
    \hline
    So I am reading the reviews and the elevator is still out. Maybe I should call the American’s with Disabilities. The elevator must be working for people with disabilities. And to say the less, the key card took forever to get activated & Maybe I should call the American’s with Disabilities. The elevator must be working for people with disabilities. & Negative \\
    \hline
    Handicap spots are not accessible during lunch, they say call but the line is busy or rings and rings. Other than that great food. & Handicap spots are not accessible during lunch, they say call but the line is busy or rings and rings. & Negative \\
    \hline
    Closet was very small and difficult to close once you hung clothes on hangers in it. Seems like we might have had a handicap room and that could be the reason for the unusual setup of the closet and the bathroom. & Seems like we might have had a handicap room and that could be the reason for the unusual setup of the closet and the bathroom. & Neutral \\
    \hline
    Very nice place to taste good oriental food. At prices very accessible to all people ... & At prices very accessible to all people ... & Unrelated \\
    \hline
    I always had my morning coffee here in this very coffee shop near the mall. It was very accessible and near my workplace. I love their coffee compared to other coffee shops in the city. I love pairing my coffee with their muffins. It was like salt and pepper, a perfect match. I recommend this coffee shop to the coffee lovers out there! & It was very accessible and near my workplace. & Unrelated \\
    \hline
  \end{tabular}
  \label{tab:example}
\end{table*}

Following this way, two authors are involved in the data annotation process. Both authors have read the ADA Accessibility Guidelines and acknowledged the criteria to classify the attitude based on Google Maps reviews. We further select a randomly sampled set of 200 reviews to verify the agreement between two annotators. These sampled reviews are labeled independently by each annotator. We use Krippendorff's $alpha$ ~\cite{krippendorff2018content} to measure the inter-coder agreement, as described by the following equation.

\begin{equation}
    \alpha = 1 - \frac{D_o}{D_e}
\end{equation}

Where:
\begin{itemize}
    \item $D_o$ is the observed disagreement between annotators
    \item $D_e$ is the expected disagreement by chance
\end{itemize}

The inter-coder agreement for our annotation, measured by Krippendorff's $alpha$, is 0.87, indicating a strong level of agreement. Following this, two annotators label a total of 2,840 randomly sampled reviews, dividing them into 2,272 for training and 568 for testing, based on the 80/20 split principle. Consequently, the training set comprises 531 negative, 103 neutral, 458 positive, and 1,180 unrelated reviews, while the testing set consists of  127 negative, 23 neutral, 129 positive, and 289 unrelated reviews.

\subsection{Attitude classification}

\subsubsection{Sentiment models}

As prior studies have used sentiment analysis to gauge public attitudes, we start by testing a sentiment analysis model called RoBERTa Sentiment ~\cite{barbieri-etal-2020-tweeteval, loureiro-etal-2022-timelms}. It is trained based on RoBERTa framework, which is a robustly BERT pre-training approach. The RoBERTa sentiment model used in this study is trained on a corpus of 123.86 million tweets collected up to the end of 2021 and fine-tuned for sentiment analysis based on the TweetEval benchmark.

\subsubsection{TF-IDF baseline models}

With the training and testing datasets prepared, the subsequent phase involves constructing and assessing classification models (Figure \ref{fig:framework}c). We incorporate three conventional NLP text classification models utilizing Term Frequency-Inverse Document Frequency (TF-IDF). TF-IDF is a measure that assesses the significance of a word within a document in relation to a set of documents (corpus). The formula for calculating TF-IDF is as follows:

\begin{equation}
    \text{TF-IDF}(t, d) = \text{TF}(t, d) \times \text{IDF}(t)
\end{equation}

Where:
\begin{itemize}
    \item $\text{TF}(t, d)$ is the frequency of term $t$ in review $d$ divided by the total number of terms in $d$.
    \item $\text{IDF}(t)$ is the inverse document frequency of term $t$, calculated as:
    \[
    \text{IDF}(t) = \log \left( \frac{N}{1 + \text{DF}(t)} \right)
    \]
    Where $N$ is the total number of Google Maps reviews in the corpus, and $\text{DF}(t)$ is the number of reviews containing the term $t$.
\end{itemize}

It is important to note that TF-IDF does not capture the semantic meaning of terms and therefore cannot consider the context in which a word appears in a review. Next, we integrate TF-IDF with three machine learning classifiers: random forest (RF), stochastic gradient descent (SGD), and logistic regression (LR). This results in three baseline models: TF-IDF+RF, TF-IDF+SGD, and TF-IDF+LR.

We conduct hyperparameter tuning on the training dataset using K-fold cross-validation with $K=10$ for each of the three TF-IDF baseline models. Table \ref{tab:grid_search} displays the hyperparameters and their corresponding grid search values for each model. During the training iteration, the model is trained on nine subsets and validated on the remaining subset. By employing 10-fold cross-validation, we evaluate the listed hyperparameter values and choose the optimal one for each model.

\subsubsection{Large language models (LLMs)}

Additionally, we utilize two LLMs for the classification task: BERT\cite{devlin-etal-2019-bert} and Llama 3\cite{dubey2024llama}. For BERT, we use the bert-base-uncased model with 110M parameters. Then, we fine-tune the BERT model to enhance its performance on accessibility-related sentiment classification. The fine-tuning process involves K-fold cross-validation with $K=5$. We also perform a grid search to optimize key hyperparameters including the batch size and number of epochs. The grid search is listed in Table \ref{tab:grid_search}.

For Llama 3, we select the Llama-3-8B-Instruct as the backbone model, taking into account the available computational resources. We fine-tune the Llama 3 model to generate the labels using the LoRA \cite{hu2021lora} technique for 10 epochs. LoRA is an efficient fine-tuning method that introduces low-rank updates to model parameters, reducing the memory and computational overhead during training \cite{hu2021lora}. The LoRA rank is set to 32, with a scaling factor $\alpha$ of 16, leading to a total of 83.89M training parameters. To stabilize training, we use a learning rate of 3e-5 and a batch size of 64. To retain the instruction-following capabilities of Llama-3-8B-Instruct on our sentiment analysis task, we also design a specific system prompt (shown in the box below) and pre-process it with data samples following the standard OpenAI chat format.


\begin{tcolorbox}[colback=gray!10!white, colframe=gray!50!gray, halign=left, top=0mm, bottom=0mm, left=1mm, right=1mm, boxrule=0.5pt]
\fontsize{9pt}{9pt}\selectfont 
\linespread{1}\selectfont
\textit{You are a language model trained to identify the sentiment for the accessibility from reviews. Your task is to analyze the text of each review and assign an appropriate label based on the sentiment or relevance of the content. The possible labels are:\\
\begin{itemize}
    \item Negative: The review expresses a negative sentiment or criticism about the accessibility of disabilities.
    \item Positive: The review expresses a positive sentiment or praise about the accessibility for disabilities.
    \item Neutral: The review provides a factual or mixed description without strong positive or negative sentiment.
    \item Unrelated: The review does not pertain to accessibility or describe any attitude toward accessibility. Reviews that discuss personal experiences or talk about something or somewhere accessible but not for disabilities, even have positive or negative words, should be labeled as unrelated.
\end{itemize}
For each input, respond with the label that best describes the sentiment or relevance of the review.}
\end{tcolorbox}

\begin{table}[htbp]
  \centering
  \caption{. Grid search values for candidate models}
  \begin{tabular}{p{0.12\linewidth}p{0.3\linewidth}p{0.45\linewidth}}
    \hline
    \textbf{Model} & \textbf{Hyperparameter} & \textbf{Grid search} \\
    \hline
    TF-IDF+RF &
    \begin{tabular}[t]{@{}l@{}}
      min\_samples\_leaf \\
      n\_estimators \\
      max\_depth
    \end{tabular} &
    \begin{tabular}[t]{@{}l@{}}
      1, 2, 4 \\
      100, 200, 300, 400 \\
      10, 20, 40, 80, 100, 120
    \end{tabular} \\
    \hline
    TF-IDF+SGD &
    \begin{tabular}[t]{@{}l@{}}
      clf\_\_alpha \\
      clf\_\_max\_iter \\
      clf\_\_penalty
    \end{tabular} &
    \begin{tabular}[t]{@{}l@{}}
      1e-4, 1e-3, 1e-2, 1e-1, 1, 10 \\
      500, 800, 1000, 2000, 3000 \\
      l2, l1, elasticnet
    \end{tabular} \\
    \hline
    TF-IDF+LR &
    \begin{tabular}[t]{@{}l@{}}
      clf\_\_C \\
      clf\_\_max\_iter \\
      clf\_\_solver
    \end{tabular} &
    \begin{tabular}[t]{@{}l@{}}
      0.1, 0.5, 1, 2, 5, 10, 20 \\
      10, 20, 50, 100, 200 \\
      sag, saga, lbfgs, newton-cg
    \end{tabular} \\
    \hline
    BERT &
    \begin{tabular}[t]{@{}l@{}}
      num\_epochs \\
      batch\_size \\
    \end{tabular} &
    \begin{tabular}[t]{@{}l@{}}
      1, 2, 3, 4, 5 \\
      16, 32, 64 \\
    \end{tabular} \\
    \hline
    Llama 3 &
    \begin{tabular}[t]{@{}l@{}}
      num\_epochs \\
      learning\_rate \\
      batch\_size \\
    \end{tabular} &
    \begin{tabular}[t]{@{}l@{}}
      5, 6, 7, 8, 9, 10 \\
      1e-5, 2e-5, 3e-5 \\
      32, 64\\
    \end{tabular} \\
    \hline
  \end{tabular}
  \label{tab:grid_search}
\end{table}

\subsubsection{Performance measures}

\begin{figure*}[htbp]
  \centering
  \includegraphics[width=0.95\textwidth]{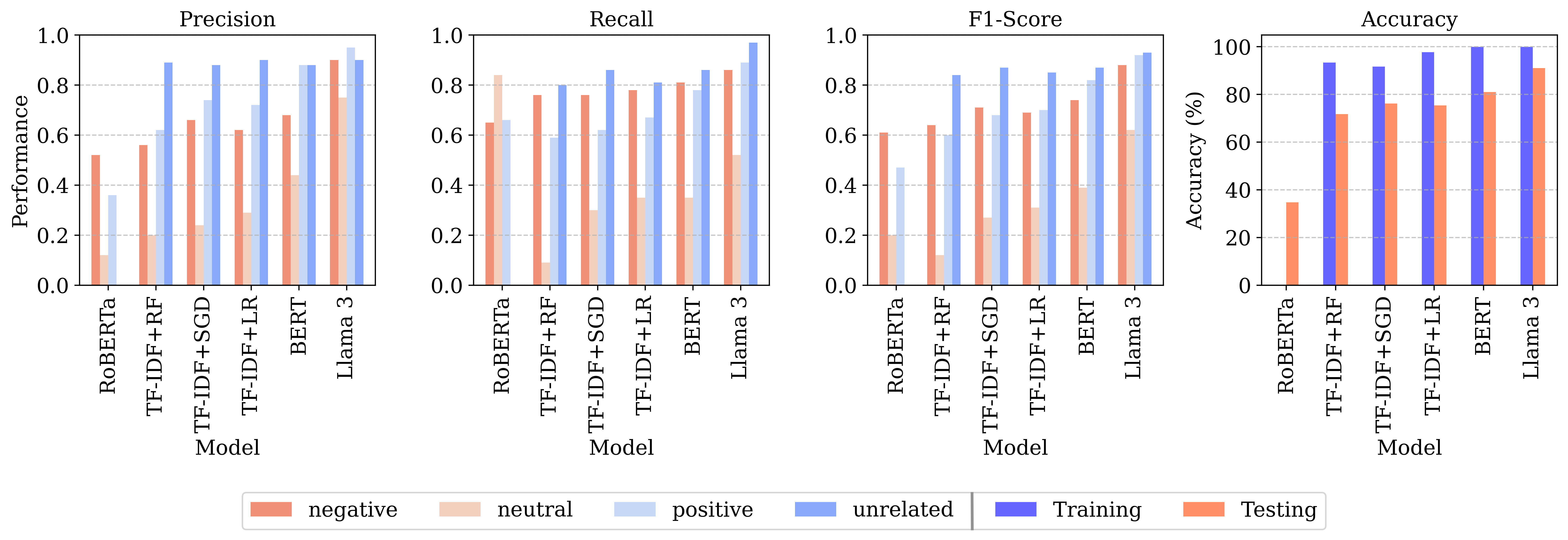}
  \caption{Classification performance of candidate models measured by Precision, Recall, F1-score, and Accuracy.}
  \label{fig:performance}
\end{figure*}

After fine-tuning the hyperparameters of baseline models and the two LLMs, we evaluate the performance of these six candidate models: RoBERTa Sentiment, TF-IDF + RF, TF-IDF + SGD, TF-IDF + LR, BERT, and Llama 3. Given that the number of samples in each category is imbalanced, the performance of these candidate models is assessed using four key metrics: Precision, Recall, F1-score, and Accuracy. Precision indicates the percentage of TP cases among all reviews predicted correctly by a respective model. Recall reflects the percentage of TP cases out of all actual positive cases. The F1-score provides a harmonic mean of precision and recall, balancing both metrics. Accuracy measures the proportion of correctly classified reviews out of all reviews. The evaluation is performed on the testing dataset, and the performance results of each model are shown in Figure \ref{fig:performance}.







Several key observations can be drawn from Figure \ref{fig:performance}. The sentiment analysis tool is not valid for our task, due to its inability to identify unrelated comments or refer to our specific contexts. TF-IDF-based models, while potentially useful for differentiating terms in reviews, struggle to interpret semantic meanings in this context, which could significantly limit their performance. The BERT model shows improvement and obtains an overall accuracy of 81\%. However, due to the imbalanced nature of the training samples, its performance in the neutral class remains limited. Last, fine-tuned Llama 3, possibly due to its large-scale pre-training process and robust text understanding capabilities, achieves the highest testing accuracy and a more balanced and higher F1-score across all classes. Based on the performance results, we select the fine-tuned Llama 3 as the best-performing model and apply it to the entire dataset for subsequent analysis.

\subsection{Semantic analysis}

With sentiment classified for each review, we then explore how individual words influence overall sentiment for different POI types. We use a tool called lexical salience-valence analysis (LSVA) \citep{taecharungroj2019analysing} to examine the words within reviews. LSVA performs text mining and evaluates the relationships between words and the sentiments expressed in the reviews by defining salience and valence. Unlike simple word frequency counts in positive or negative reviews, the LSVA method provides a visualization of word frequency across the document corpus and its impacts on overall sentiment. LSVA defines a word's salience and valence in the following way:

\begin{equation}
    \centering
    \text{salience}|_{\text{word}_i} = \log_{10}(N_{\text{total}})|_{\text{word}_i}
    \label{euqation: salience}
\end{equation}

\begin{equation}
    \centering
    \text{valence}|_{\text{word}_i} = \frac{N(\text{positive}) - N(\text{negative})}{N_{\text{total}}}|_{\text{word}_i}
    \label{euqation: valence}
\end{equation}

Where:
\begin{itemize}
    \item $N_{\text{total}}$ represents the number of reviews in which the term $word_i$ appears.
    \item $N(\text{positive})$ represents the number of positive reviews in which $word_i$ appears.
    \item $N(\text{negative})$ represents the number of negative reviews in which $word_i$ appears.
    \item The salience of a term is computed by the logarithm with a base 10 function of the frequency of each term.
    \item The valence of a term is computed as $N(\text{positive}) - N(\text{negative})$ divided by its total count $N_{\text{total}}$, which measures how positive a particular term appears in a corpus.
\end{itemize}

\subsection{Regression modeling}
To assess whether public perceptions of accessibility vary across regions with different socioeconomic statuses, we construct a regression model correlating accessibility sentiment with nearby socio-spatial features. We calculate the average sentiment from all accessibility-related reviews within each region, selecting regions with more than 10 reviews as our regression samples. Counties with insufficient reviews are excluded because the limited number of reviews cannot reliably represent the overall sentiment of the entire county. Sensitivity analysis is then performed by adjusting the minimum review threshold from 0 to 50. Results show that regression results stabilize after a threshold of 10. To address the modifiable area unit problem (MAUP) \cite{hu2022examining}, which suggests that outcomes can vary based on spatial aggregation levels, we fit the regression at both county and census block group (CBG) levels. The regression is fitted under the generalized additive model (GAM) framework. The GAM is a semi-parametric framework that combines parametric linear predictors with a suite of additive, non-parametric nonlinear predictors. These nonlinear components are fitted using various types of smooth splines, enhancing the model's flexibility and adaptability. In our specification, the effects of main covariates of interest are modeled with linear terms, whereas the geospatial correlation between counties/CBGs that are located closer to each other is modeled with a smoother function. We additionally allow for state-specific random effects. Therefore, our regression model is formulated as: 

\begin{equation}
g(E(Y)) = \beta_0 + \sum_{k=1}^{N_l} \beta_k X_k + ti(\text{SR}) + s(\text{RE}) + \epsilon
\end{equation}

Where:

\begin{itemize}
    \item \(Y\) is CBG- or county-level sentiment, assuming that both follow the Gaussian distribution.
    \item \(g(\cdot)\) refers to the link function, which is the identity function in this case.
    \item \(\beta_0\) is the overall linear intercept.
    \item \(\beta_k\) is the coefficient of the \textit{kth} variable \(X_k\), where \(X_k\) is the variable set with linear effects. Here it includes all socio-spatial factors outlined in Table \ref{tab:variable_summary}.
    \item \(SR\) denotes the spatial autocorrelation term. Here spatial autocorrelation is fitted by \(ti(\cdot)\), which is a marginal nonlinear smoother that excludes main effects.
    \item \(RE\) denotes the state-level random effects captured by the penalized spline function \(s(\cdot)\).
    \item \(\epsilon\) is the error term.
\end{itemize}

The data sources for the independent variables are listed as below: Socioeconomics and demographics are from the 2018-2022 5-year American Community Survey (ACS) \cite{ACS2022}; POI variables are from Google Maps; Road density is collected from OpenStreetMap. Comprehensive details and descriptive statistics for all independent variables (CBG-level) are presented in Table \ref{tab:variable_summary}. Variables in \textit{italic} are excluded from the models due to high multicollinearity, indicated by a variance inflation factor (VIF) greater than 5. Additionally, to mitigate the effects of outliers, influential observations are removed if their Cook's distance exceeds $\textit{4/(n - k - 1)}$, where \textit{n} is the regression sample size and \textit{k} is the total number of independent variables in the regression \cite{belsley2005regression}.

\begin{table*}[]
\centering
\caption{. Summary of independent variables in the regression model}
\label{tab:variable_summary}
\resizebox{0.95\linewidth}{!}{%
    \begin{tabular}{p{0.2\linewidth}p{0.5\linewidth}p{0.1\linewidth}p{0.1\linewidth}}
        \hline
        \textbf{Variable} & \textbf{Description} & \textbf{Mean} & \textbf{Std.} \\ \hline
        \textbf{Socioeconomic} &  &  &  \\ \hline
        Population Density & Population density (people per acre) & 4.822 & 9.577 \\ \hline
        Employment Density & Employment density (jobs per acre) & 11.015 & 26.342 \\ \hline
        Poverty & \% households below poverty & 14.179 & 11.888 \\ \hline
        Rural Population & \% rural population & 8.399 & 22.887 \\ \hline
        \textit{Urban Population} & \% \textit{urbanized population} & 83.162 & 35.362 \\ \hline
        \textit{Median Income} & \textit{Median household income (Inflation-Adjusted), in \$ $10^{3}$/household} & 6.419 & 3.065 \\ \hline
        Highly-Educated & \% residents with at least a Bachelor's degree & 35.146 & 19.993 \\ \hline
        \textbf{Demographics} &  &  &  \\ \hline
        Male & \% males & 49.332 & 6.812 \\ \hline
        Age 18-44 & \% residents aged 18 to 44 years & 39.167 & 15.004 \\ \hline
        Age 45-64 & \% residents aged 45 to 64 years & 24.721 & 8.249 \\ \hline
        Age over 65 & \% residents aged 65 and older & 17.475 & 11.907 \\ \hline
        \textit{White} & \textit{\% Non-Hispanic Whites }& 62.597 & 25.470 \\ \hline
        Asian & \% Asians & 5.904 & 9.557 \\ \hline
        African American & \% African Americans & 11.252 & 16.244 \\ \hline
        Hispanic & \% Hispanics/Latinos & 16.497 & 19.160 \\ \hline
        Others & \% Other minorities. & 0.936 & 4.093 \\ \hline
        Disability & \% residents with a disability. & 11.623  & 11.434  \\ \hline
        \textbf{Review} &  &  &  \\ \hline
        Avg. POI Score & Average POI score for all POIs collected from Google Maps, regardless of POI types or review contents & 4.250 & 0.182 \\ \hline
        Review density & Total accessibility-related review density & 0.393 & 1.111 \\ \hline
    \end{tabular}%
}
\end{table*}

\section{Results}

The result analysis explores two research questions. For the first research question, we examine whether the public perception of accessibility varies across different POI types. Specifically, we investigate two aspects: whether different POI types exhibit variation in public sentiment, and what factors or aspects extracted from reviews contribute to the sentiment for different POI types. For the second research question, we explore the variation in public perception of accessibility from the geospatial perspective. Again, we divide this into two parts: identifying the patterns in public perception across different geographic areas, and examining the important social, demographic, and economic factors associated with the public sentiments.

Each POI in Google Maps is associated with a category. This category can be mapped onto the North American Industry Classification System (NAICS) codes \cite{naics}. For example, a POI labeled as ``restaurant'' by Google Maps corresponds to the 3-digit NAICS code ``722-Food Services and Drinking Places,'' while a ``Grocery store'' corresponds to ``445-Food and Beverage Stores.'' However, with 102 3-digit NAICS codes, interpretation can present challenges. To address this, we manually group frequently mentioned Google Maps categories into several major POI types that are often associated with high-volume daily activities. Other POI types such as manufacturing facilities, production plants, or construction companies are not included in subsequent analysis as they are not closely related to the public's daily activities. For the subsequent analysis, the POI types we focus on are:

\begin{itemize}
    \item Restaurant: Food and drink providers, e.g., restaurants, bars, drinking places.
    \item Retail Trade: Merchandise sellers, e.g., grocery stores, convenience stores, home goods stores. 
    \item Recreation: Leisure and entertainment facilities, e.g., recreation centers, stadiums, gyms, playgrounds. 
    \item Hotel: Short-term lodging providers, e.g., hotels, motels. 
    \item Personal Service: Consumer-oriented service businesses, e.g., spas, salons, massage therapists. 
    \item Health Care: Medical facilities and services, e.g., medical centers, urgent cares, dentists, hospitals. 
    \item Transportation: People and goods movement services, e.g., bus stops, bus stations, shipping services. 
    \item Public Service: Government and community services, e.g., business centers, mailbox, security services, locksmiths. 
    \item Apartment: Multi-unit residential buildings, e.g., apartments.
    \item Education: Learning and instruction institutions, e.g., schools, universities, colleges. 
\end{itemize}


\subsection{POI analysis}

\subsubsection{What are the sentiment patterns among POI types?}
\begin{figure*}[htbp]
  \centering
  \includegraphics[width=0.95\textwidth]{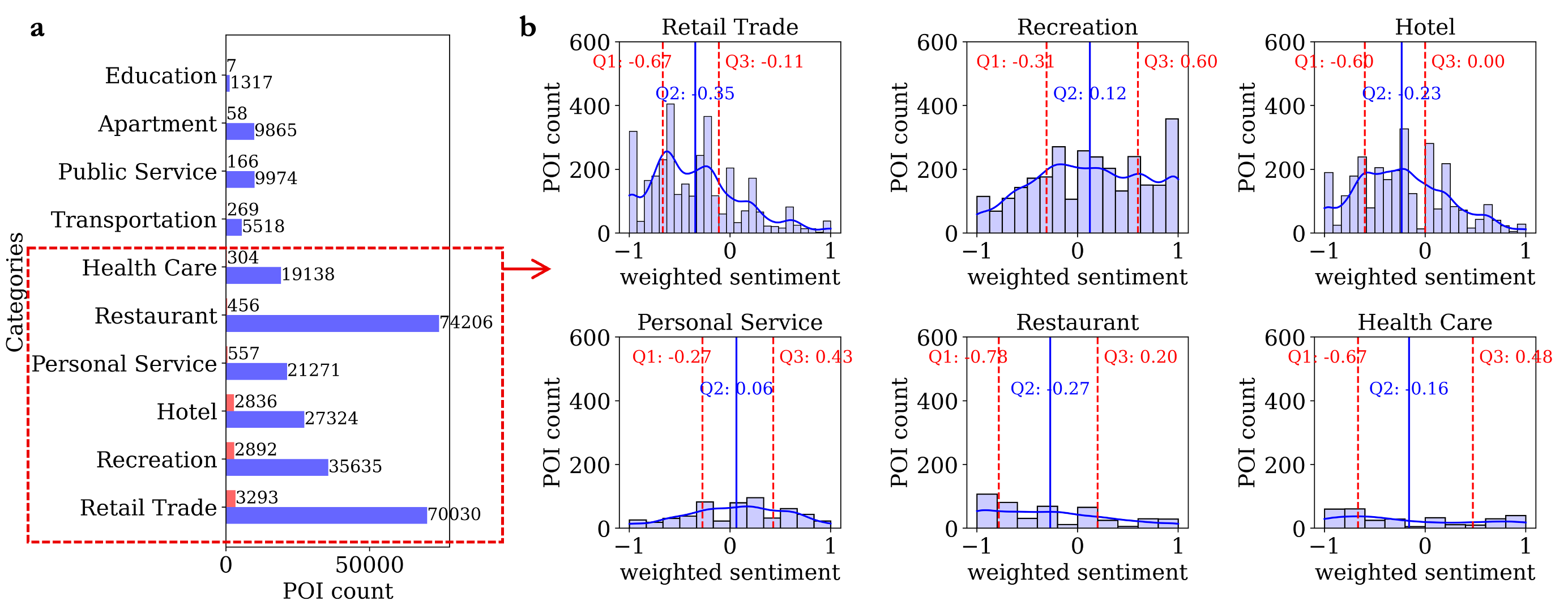}
  \caption{Descriptive results for sentiment by POI. (a) Number of POIs by types. (b) Distribution of sentiment. The x-axis represents the weighted sentiment by averaging all sentiments associated with a POI, while the y-axis shows the POI count.}
  \label{fig:poi_distribution}
\end{figure*}

To address RQ1(1), we begin by presenting the descriptive results categorized by POI types, as illustrated in Figure~\ref{fig:poi_distribution}. When considering all reviews for analysis (the blue bar in Figure~\ref{fig:poi_distribution}(a)),  the most commonly mentioned POI types include Restaurant, Retail Trade, Recreation, Hotel, Personal Service, and Health Care. However, when applying a threshold of five reviews or more (the red bar in Figure~\ref{fig:poi_distribution}(a)), the order changes slightly, with Retail Trade, Recreation, Hotel, Personal Service, Restaurant, and Health Care being the most mentioned, in which Restaurants show a significantly smaller proportion of POIs with five or more reviews. This suggests that while restaurants and drinking places are the most commonly reviewed POIs, most of them receive a very small number of comments that mention accessibility.

We select six major POI types for subsequent analysis given the following two reasons. First, they are closely associated with daily activities that produce high visit volumes, accounting for over 90\% of total reviews across all POIs. Second, local businesses such as restaurants, retail trades, and recreation centers often experience high turnover rates and busy demand periods, indicating that accessibility issues could be more pronounced at these POIs. In addition, to avoid small sample bias from POIs with only one or two comments, we select those POIs with at least five reviews mentioning accessibility to plot the sentiment distribution. The sentiment distribution for each of these POI types is illustrated in Figure~\ref{fig:poi_distribution}(b).

The sentiment analysis across different POI types reveals diverse patterns and insights (Figure~\ref{fig:poi_distribution}(b)). Most POI categories exhibit negative average sentiments, with Retail Trade showing the most negative outlook with all three quartiles being negative (Q1: -0.67, Q2: -0.35, and Q3: -0.11). Recreation is the only POI type with a positive average sentiment (Q2: 0.12). Restaurant presents the highest negative first quartile (Q1: -0.78).

In addition, the distribution patterns offer further insights into sector-specific experiences. Retail Trade and Restaurant display right-skewed distributions indicating a tendency toward negative experiences with accessibility. Personal Service and Hotel show approximately normal distributions. In particular, Personal Service shows a relatively normal distribution centered near neutral, implying balanced experiences. Moreover, Recreation's left-skewed distribution aligns with its overall positive sentiment. Health Care exhibits a unique ``U-shaped'' distribution, with higher frequencies at both extremes (-1 and 1), suggesting polarized opinions about healthcare accessibility.

\subsubsection{What are the aspects explaining the sentiments among POI types?}

\begin{figure*}[htbp]
  \centering
  \includegraphics[width=0.95\textwidth]{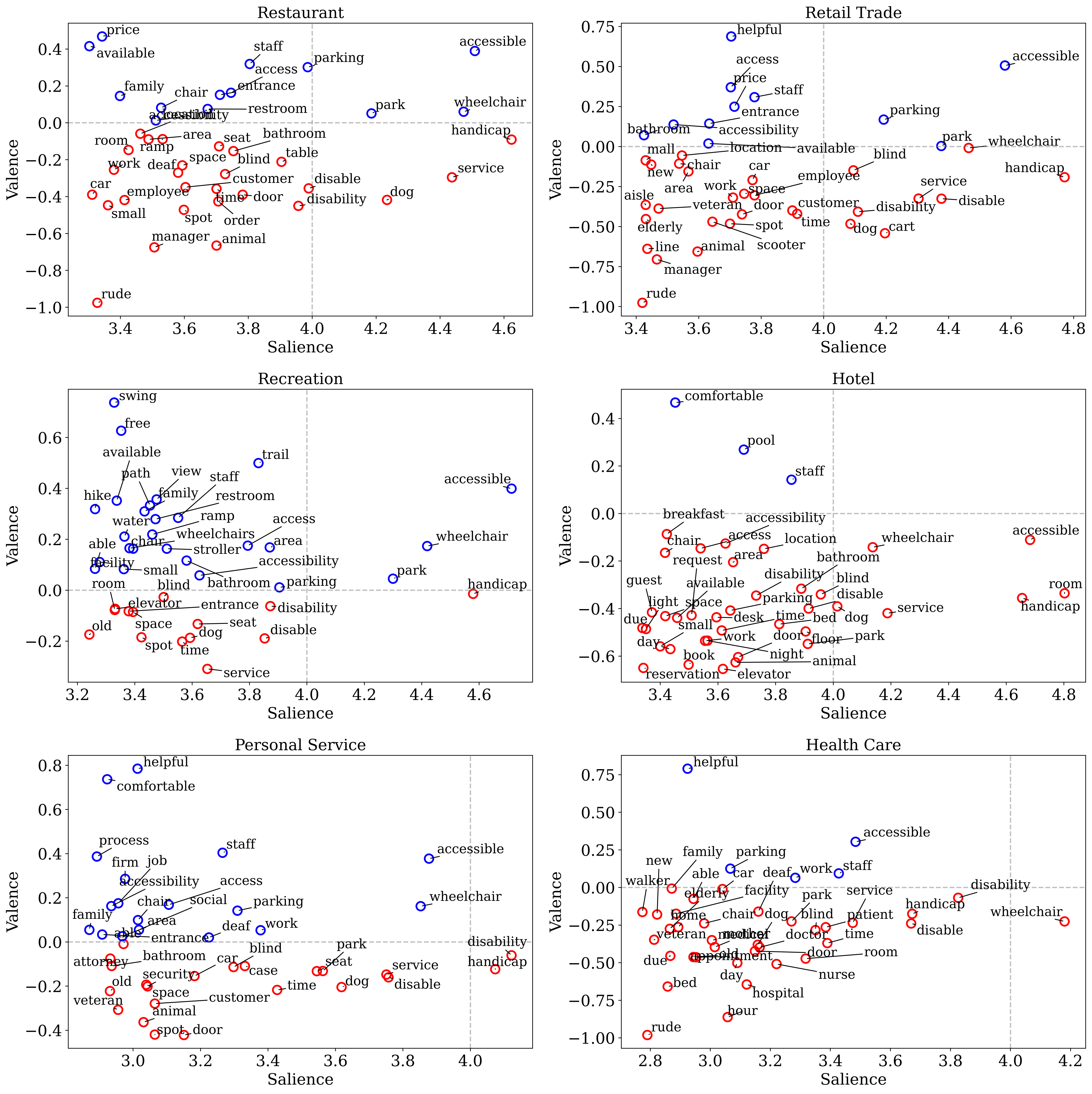}
  \caption{Semantic analysis using LSVA based on POI types. The x-axis represents the salience using equation \ref{euqation: salience}, while the y-axis represents the valence using equation \ref{euqation: valence}.}
  \label{fig:text}
\end{figure*}

In response to RQ1(2), we delve into the aspects that explain the sentiments across the six major POI types, including Restaurant, Retail, Recreation, Hotel, Personal Service, and Health Care. We follow equations~\ref{euqation: salience} and ~\ref{euqation: valence} to calculate the salience and valence of each term in the corpus of reviews in terms of each POI type. Figure~\ref{fig:text} shows the unique accessibility-related sentiments for each POI type, facilitating better understanding by exploring the aspects contributing to positive and negative experiences, respectively marked in red and blue circles. 

For Restaurant, negative sentiments are predominantly associated with staff interactions, service, the presence of service animals, particularly service dogs, and barriers such as steps and ramps. Accessibility-related terms such as ``wheelchair,'' ``entrance,'' and ``parking'' appear with  neutral or slightly positive sentiments. This indicates that while structural accessibility features may be in place, social factors such as service and staff interactions contribute to negative sentiment.

In the Retail sector, accessible features such as ``parking,'' ``entrance,'' and ``wheelchair'' appear on the positive side of the valence axis, indicating appreciation for outdoor physical accommodations. However, negative sentiments stem from issues like ``door,'' ``employee,'' ``line,'' ``aisle,'' and ``cart,'' which suggest challenges with the indoor environment in stores, such as handling accessibility for customers who require physical accommodations as well as customer service issues.

For Recreation, more positive sentiments appear. Those positive aspects are mainly associated with terms such as ``accessible,'' ``park,'' ``ramp,'' ``facility,''``trail,'' and ``family,'' highlighting the positive impact of inclusive environments, particularly for family-friendly spaces. On the negative side, aspects such as ``service,'' ``staff,'' and ``entrance'' appear, showing that despite the availability of accessible features, other operational issues such as poor service or design negatively impact individuals' experiences.

The Hotel category presents a more imbalanced sentiment distribution, with only a few positive aspects shown in the figure, including ``pool'' and ``staff.'' Negative sentiments dominate the terms, arising from infrastructure issues related to ``room,'' ``elevator,'' and ``bathroom,'' to service issues related to ``guest,'' ``book,'' and ``reservation.'' This highlight areas where accommodation for accessibility needs may fall short, particularly in terms of room features and staff assistance.

In the Personal Service sector, the positive aspects include ``comfortable,'' ``staff,'' and ``process,'' reflecting satisfaction with the ease of service and assistance provided. However, terms like ``seat,'' ``space,'' ``spot,'' ``handicap,'' and ``bathroom'' reflect negative experiences, suggesting challenges related to accessibility facilities.

For Health Care, the highly negative terms associated with sentiments include ``doctor,'' ``nurse,'' ``service,'' ``appointment,''  and ``time'' reflecting frustrations with service received. Other negative terms such as ``room'' and ``door'' reflect negative experiences with medical environments. However, positive aspects such as ``staff,'' ``helpful,'' and ``family'' indicate appreciation for supportive staff and accessible infrastructure in healthcare facilities.

Across all POI types, accessible physical infrastructure (e.g., parking, wheelchair access, entrance, ramp) tends to receive positive feedback, while issues with staff (e.g., employee, manager, doctor, nurse), service (e.g., service animal, reservation, book), and accommodations for individuals with disabilities (e.g., space, spot, bathroom, seat) lead to more negative sentiments. This highlights the multifaceted nature of accessibility, where both physical infrastructure and human or social factors play crucial roles in shaping accessibility experiences.

\subsection{Socio-spatial analysis}
\subsubsection{What are the sentiment patterns across geographic regions?}

\begin{figure}[htbp]
  \centering
  \includegraphics[width=0.5\textwidth]{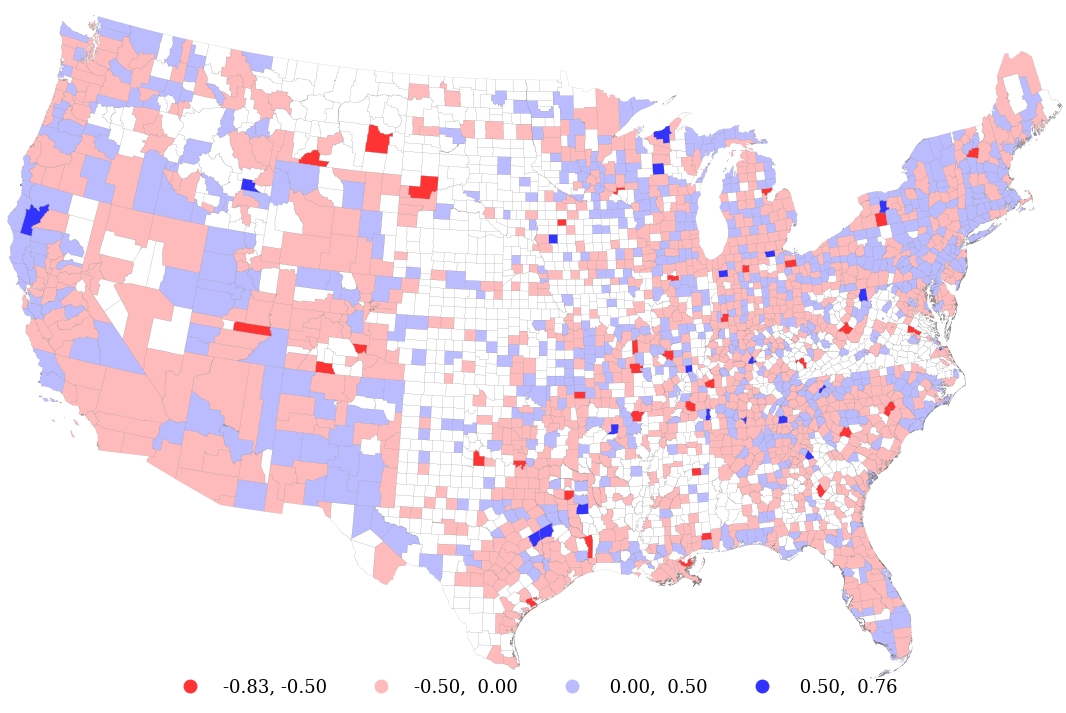}
  \caption{Spatial distribution of accessibility sentiment.}
  \label{fig:cbg_map}
\end{figure}

We plot the variation in accessibility sentiment across all counties in the contiguous United States in Figure \ref{fig:cbg_map}. Counties with negative sentiment are depicted in blue, those with positive sentiment in red, and counties with fewer than 10 reviews are left uncolored. The choropleth map underscores a high dispersion in accessibility sentiment, with positive and negative reviews appearing sporadically, lacking a clear spatial clustering pattern. This is further evidenced by the insignificant spatial interaction term in \autoref{tab:regress_outcome}. However, a notable pattern is observed in the distribution of absent reviews: counties in the Midwest regions tend to have fewer accessibility-related reviews, whereas counties in the Northeast, South Atlantic, and West are more likely to leave reviews that mention accessibility. This may be attributed to the lower population densities and public service availability in these regions.

\begin{figure}[htbp]
  \centering
  \includegraphics[width=0.5\textwidth]{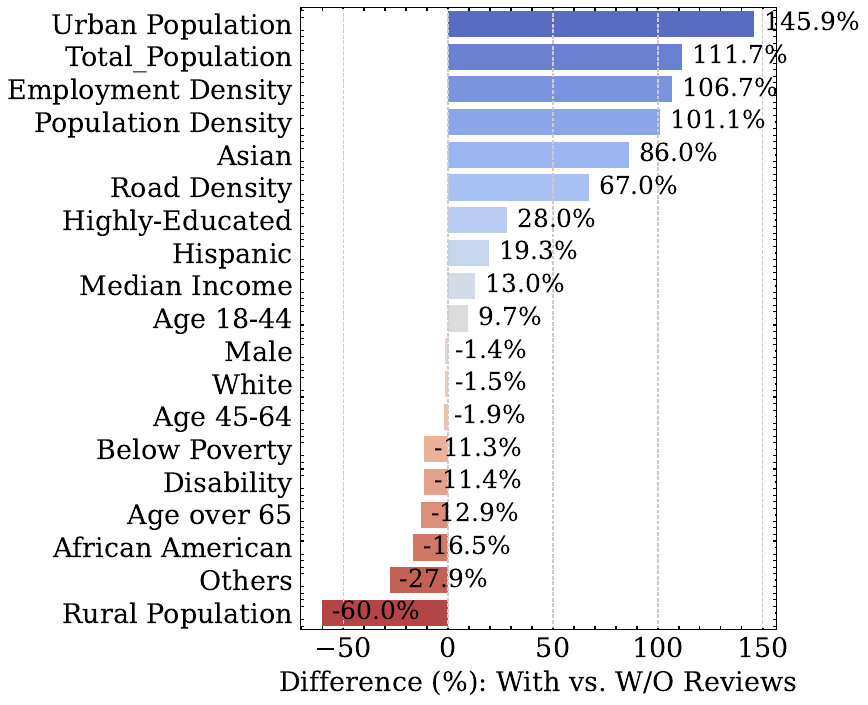}
  \caption{Socio-spatial difference in accessibility review engagement: Counties with and w/o $>$ 10 accessibility reviews.}
  \label{fig:diff_socio}
\end{figure}

We further explore the socio-spatial differences between counties based on the presence of more than 10 accessibility-related reviews. Of the total, 1,976 counties have more than 10 reviews, while 1,080 counties do not. As shown in Figure \ref{fig:diff_socio}, counties with higher population densities, greater employment densities, and higher levels of urbanization are more likely to accumulate a substantial number of accessibility reviews. Notable racial disparities also emerge, particularly among Asians, African Americans, and other racial minorities, which may reflect the location preferences associated with these groups' activities. This comparison highlights a potential limitation of the sentiment regression analysis: the counties included in regression are not randomly selected but rather show selection biases. These biases are influenced by the tendencies of individuals who post reviews on Google Maps, as well as by the more densely developed regions where reviews are more frequently collected.

\subsubsection{How are sentiment patterns associated with local socio-spatial factors?}

We first perform a univariate correlation analysis at the county level, identifying the top three positively correlated features as: Avg. POI Score (0.169), White (0.139), and Age 45-64 (0.098). Conversely, the top three negatively correlated features were: African American (-0.119), Hispanic (-0.077), and Poverty (-0.071). Similar patterns were observed at the CBG level, where the positive leaders were: Avg. POI Score (0.282), White (0.120), and Median Income (0.079), and the negative leaders were: Hispanic (-0.085), Poverty (-0.080), and African American (-0.077). This analysis suggests that areas with higher socioeconomic status and a larger percentage of White residents tend to have a more favorable public perception of urban accessibility. 

\begin{figure*}[!ht]
    \centering    
    \begin{subfigure}[t]{0.3\textwidth}
        \centering
        \includegraphics[width=\textwidth]{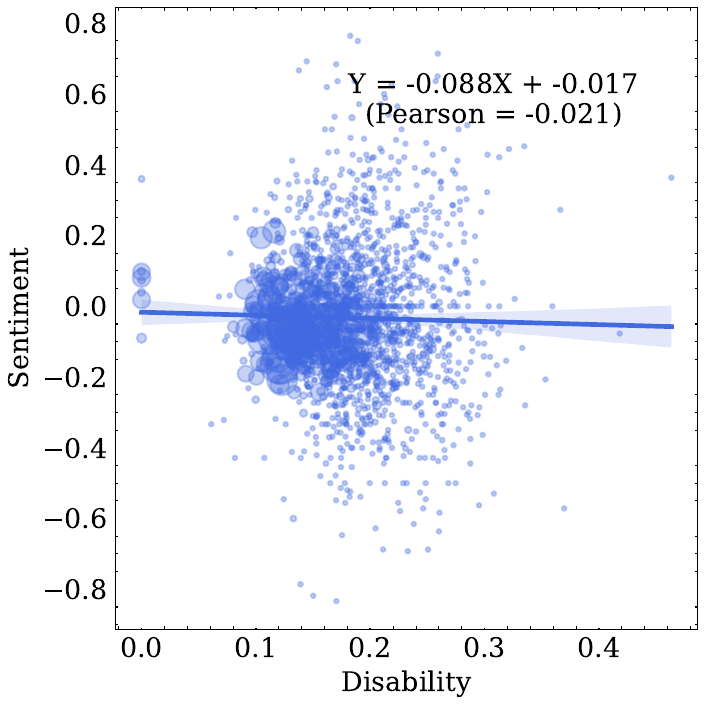}
        \caption{\% of Disability}
    \end{subfigure}%
    ~
    \begin{subfigure}[t]{0.3\textwidth}
        \centering
        \includegraphics[width=\textwidth]{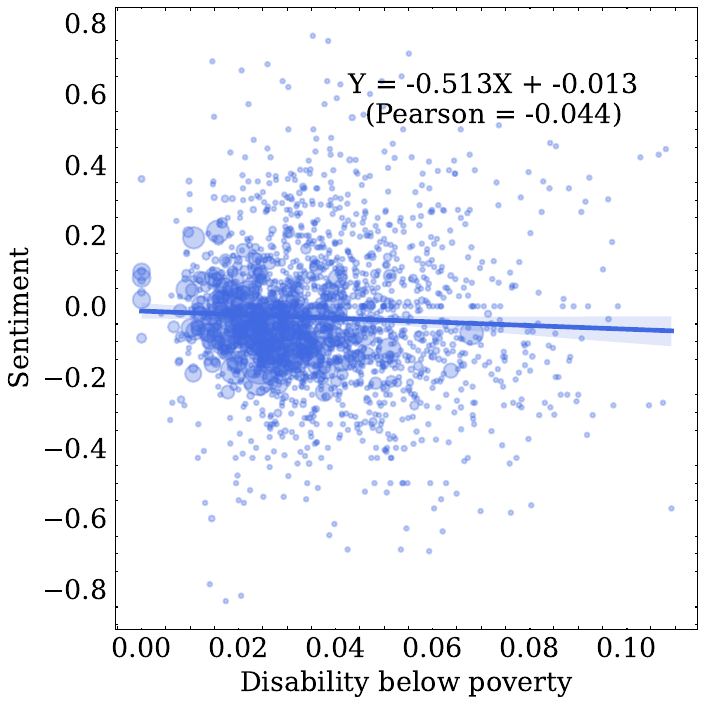}
        \caption{\% of Disability below poverty}
    \end{subfigure}%
    ~
    \begin{subfigure}[t]{0.3\textwidth}
        \centering
        \includegraphics[width=\textwidth]{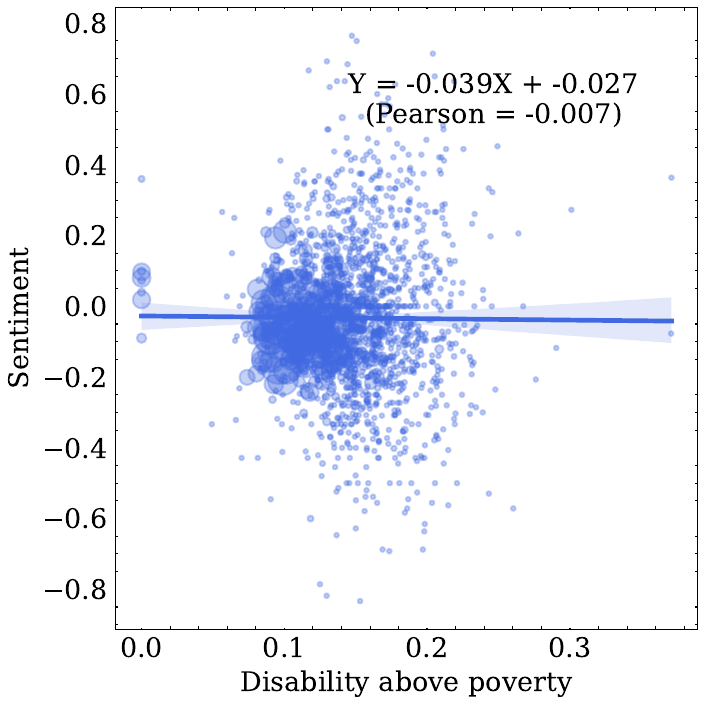}
        \caption{\% of Disability above poverty}
    \end{subfigure}%
    \caption{Scatter plot: Disability (From ACS) vs. Sentiment (From Google Map reviews)}
    \label{fig:Disability}
\end{figure*}

However, we do not find a significant correlation between accessibility sentiment and \% residents with disabilities. As detailed in \autoref{fig:Disability}, the Pearson correlation at the county level between disability and accessibility sentiment is only -0.021. Interestingly, this correlation slightly increases to -0.044 when considering only those with disabilities below poverty, and decreases to -0.007 for those above poverty. These findings highlight the underlying socioeconomic disparities in accessibility sentiment, although the links between disability and accessibility sentiment are not statistically significant.

We then fit the GAM and report the regression results in Table \ref{tab:regress_outcome}. Model goodness-of-fit, the adjusted $R^2$, is 0.126 for CBG-level modeling and 0.099 for county-level modeling. While these values are moderate, they are considered reasonable given the numerous immeasurable factors that could influence public perception. Regarding nonlinear effects, the spatial interaction terms are not statistically significant, suggesting weak spatial dependence in the spatial distribution of accessibility sentiment. However, the state-level random effects are all statistically significant, with degrees of freedom (e.d.f.) consistently greater than 1, indicating pronounced random effects at the state level. This could be attributed to unobserved effects or influences inherent to the states themselves.

The coefficients of the two models are largely similar, although there are subtle differences in their statistical significance. Demographic factors reveal that elderly populations consistently exhibit a negative relationship with accessibility sentiment. This trend may indicate the high dependence of elderly adults on barrier-free urban facilities, underscoring their unique needs within the community. Additionally,  the racial and ethnic composition of a region is significantly associated with accessibility sentiment. Areas with higher proportions of African American and Hispanic populations generally report more negative accessibility experiences compared to predominantly White regions. The percentage of males shows a significantly positive relationship with sentiment at the county level, but the significance disappears at the CBG level. Similar to the univariate correlation analysis, no clear relationship is evident between the percentage of people with disabilities and accessibility sentiment. 

Among socioeconomic factors, high employment density exhibits a strong negative correlation with accessibility sentiment but only at the CBG level. The rural population exhibits a significant negative relationship but only at the county level. Household poverty rates consistently show negative relationships with accessibility sentiment, suggesting that economically disadvantaged areas are more likely to experience subpar accessibility facilities. Interestingly, regions with more highly educated residents tend to exhibit more negative sentiments towards accessibility. This may be because these individuals are more attuned to the design and functionality of accessibility facilities and are also more likely to express their opinions through online reviews. Lastly, a strong (also the strongest among all variables) positive relationship is found between averaged POI scores and accessibility sentiment, which is plausible as the average POI score represents the overall user experience with local services in the region. 

\begin{table}[]
\centering
\caption{. Regression outcomes (standardized)}
\label{tab:regress_outcome}
\resizebox{\linewidth}{!}{%
\begin{tabular}{p{0.4\linewidth}p{0.3\linewidth}p{0.3\linewidth}}
    \hline
    \multicolumn{1}{c}{\textbf{Variables}} & \multicolumn{1}{c}{\textbf{CBG-level}} & \textbf{County-level} \\ \hline
    (Intercept)        & -0.093***  & -0.037*** \\ \hline
    \multicolumn{3}{l}{\textbf{Demographics}}            \\ \hline
    Asian              & 0.002      & 0.001     \\ \hline
    African American   & -0.009**   & -0.021*** \\ \hline
    Hispanic           & -0.020***  & -0.020*** \\ \hline
    Others             & -0.010*    & 0.003     \\ \hline
    Male               & -0.003     & 0.015**   \\ \hline
    Age over 65        & -0.010**   & -0.017**  \\ \hline
    Age 45-64          & -0.002     & -0.005    \\ \hline
    Disability         & 0.001      & -0.001    \\ \hline
    \multicolumn{3}{l}{\textbf{Socioeconomics}}          \\ \hline
    Population Density & 0.005      & 0.025     \\ \hline
    Employment Density & -0.029***  & -0.015    \\ \hline
    Rural Population   & -0.002     & -0.022**  \\ \hline
    Poverty            & -0.011**   & -0.008**  \\ \hline
    Highly-Educated    & -0.019***  & -0.008*   \\ \hline
    \multicolumn{3}{l}{\textbf{Reviews}}                 \\ \hline
    Review density     & -0.009     & -0.017    \\ \hline
    Avg. POI score     & 0.097***   & 0.028***  \\ \hline
    \multicolumn{3}{l}{\textbf{Nonlinear terms (degrees of freedom (e.d.f.))}}         \\ \hline
    ti(Lat,Lng)        & 2.602      & 1.705     \\ \hline
    s(State)            & 128.347*** & 16.053*** \\ \hline
    \multicolumn{3}{l}{\textbf{Model fit}}               \\ \hline
    Adjusted. R2       & 0.126      & 0.099     \\ \hline
    R2                 & 0.141      & 0.116     \\ \hline
    Sample size        & 8702   & 1901  \\ \hline
    \multicolumn{3}{l}{\begin{tabular}[c]{@{}l@{}}Significance codes: 0 ‘***’ 0.001 ‘**’ 0.01 ‘*’ 0.05 ‘’ 1.\end{tabular}}
\end{tabular}%
}
\end{table}

\section{Discussion}

In this study, we collect Google Maps reviews across the United States and use fine-tuned LLMs to analyze public sentiment on accessibility, exploring the relationship between these sentiments and socio-spatial factors. The key findings are outlined below.

\textbf{First, most POI types show negative sentiments, except for Recreation.} This suggests an overall negative attitude from the public towards accessibility features in most POI types. Furthermore, sentiment analysis reveals distinct distribution patterns during the study period. For example, Retail Trade and Restaurants display right-skewed distributions, indicating a need for more attention to accessibility improvements, while Health Care shows a polarized distribution, meaning people have either very positive or very negative experiences, with little middle ground.

\textbf{Second, public perceptions of positive and negative aspects differ depending on the type of POI.} Infrastructure-related aspects like ``parking,'' ``wheelchair,'' and ``entrance'' receive positive feedback, indicating that accessibility design is generally effective in these areas. However, negative sentiments converge to human-related or social factors. Specifically, Hotels and Health Care received extensive negative feedback on aspects such as ``service,'' ``doctor,'' ``nurse,'' ``time,'' and ``room,'' emphasizing issues related to service providers and the quality of service itself. Prior research has also highlighted the challenges staff faces in communicating with guests with disabilities during service encounters in hospitality settings \cite{kalargyrou2018impact}. Although Recreation and Personal Services receive relatively fewer negative comments, their accommodation for accessibility, such as ``spot,'' ``handicap,'' and their inclusive environment for ``dog'' and ``old,'' need improvement, as these terms show negative sentiment. 

\textbf{Third, there are significant geographical disparities in sentiment toward accessibility.} Areas with higher socioeconomic status and predominantly White populations display more positive perceptions, likely due to better-funded infrastructure and services in these regions. In contrast, regions with higher Hispanic and African American populations, as well as those with higher poverty rates, show more negative sentiments, reflecting systemic neglect and fewer resources available for accessibility improvements. Additionally, areas with larger populations of elderly and highly-educated residents display strong negative sentiments toward urban accessibility, indicating these groups can be more sensitive to the design and functionality of accessibility facilities. Our study also aligns with existing findings regarding barriers to health care for individuals with disabilities who are members of underserved racial and ethnic groups \cite{peterson2014barriers}. 


\textbf{Last, no significant link is found between the presence of people with disabilities and public perceptions.} While higher densities of disabled individuals could typically suggest increased feedback, as these communities are most affected by accessibility shortcomings, the lack of a significant correlation implies that public sentiment is not necessarily reflecting their needs. This disconnect suggests that areas with larger disabled populations may not receive the necessary attention or improvements, highlighting a gap in inclusive planning. These results align with existing research, which shows that public agencies address only minimum accessibility standards and overlook the lived experiences of disabled individuals \cite{levine2023approaching}. Moreover, the lack of attention to accessibility issues in transportation planning \cite{levine2023approaching} and public services \cite{krahn2015persons} reflects a broader disconnect between policy and the actual needs of disabled communities.

\subsection{Practical implications}

The findings from this study reveal disparities in public perceptions of accessibility and their correlations with various socio-spatial factors. These findings offer critical insights for multiple stakeholders, including government agencies, local businesses, and people with disabilities and their caregivers. 

Government agencies play a pivotal role in shaping policies and infrastructure to enhance public spaces. Our findings highlight two critical areas for improvement. First, significant concerns regarding staff behavior and the adequacy of services and accommodations suggest a need for government agencies to offer staff training programs. The training should ensure that staff are adequately prepared to accommodate people with disabilities. Second, the geographic disparities identified in our study, particularly the negative correlation between accessibility and factors such as poverty, racial and ethnic composition, and employment density, further recognize the necessity for equitable distribution of resources. Public investments in accessible infrastructure should target underrepresented and underserved areas, particularly those with high concentrations of low-income, minority, and elderly populations. Incentives such as subsidies or training programs should be provided to businesses in these areas to prioritize accessibility improvements.

For local businesses, improving accessibility and customer service for people with physical disabilities can serve as a competitive advantage. While physical accessibility features such as parking and entrances are often positively received, businesses need to move beyond merely complying with physical accessibility requirements and work on providing inclusive customer service. This includes training staff to be more knowledgeable and responsive to the needs of people with physical disabilities. In addition, businesses should consider adopting practices from POI types that are viewed more positively, such as recreation, to their operations. 

Furthermore, our findings demonstrate the importance of advocacy by and for people with disabilities. There are critical needs for improvements in underlying services, such as accommodations (e.g., bathroom accessibility or seating) and service staff interactions, beyond the provision of common, public-aware physical accessibility infrastructure. Individuals with disability and their caregivers should actively post on social media and participate in dialogue with businesses and policymakers to ensure that accessibility efforts extend beyond physical infrastructure to include customer service and accommodation of needs. Voices from them are crucial in raising awareness of issues that are often overlooked by the general public but are equally important to create accessible and inclusive urban environments.

\subsection{Future work}

This study presents several avenues for future research. The first opportunity lies in model classification, where improvements can be made in two key areas. First, we could engage more annotators with domain expertise to label a larger dataset for both training and testing. This would allow the model to capture a broader variety of textual information from online reviews, which could ultimately enhance the model’s robustness. Second, due to computational limitations, we only fine-tune the Llama 3-8B model in this project. Although the fine-tuned Llama model achieves 91\% accuracy, using LLMs with more parameters could potentially improve this accuracy. Future research could also explore testing closed-source models like GPT-4 by OpenAI or Gemini by Google.

The second research direction involves data fusion, which could be approached in two ways. First, our findings reveal that many POIs on Google Maps have only one or two comments related to accessibility, leading to potential sampling biases in large-scale analyses. To mitigate this, future work could incorporate reviews from other platforms like Yelp or TripAdvisor, thereby expanding the dataset and offering a more comprehensive perspective. Additionally, these insights could be integrated with data from active crowdsourcing efforts. For example, previous research has developed platforms where participants actively rate accessibility ~\cite{crowdsourcingGoogleStreetView, saha2019project}. By combining data from both active and passive crowdsourcing, we can further highlight the strengths of crowdsourcing approaches in studying accessibility in urban environments.

Next, it is important to acknowledge potential biases among users who choose to leave online reviews. Previous studies have demonstrated that social media, including online review platforms, could overrepresent certain demographics, such as male, educated, and urban populations \cite{mellon2017twitter, wang2019demographic}. While our passive crowdsourcing approach addresses some limitations of traditional surveys, it may introduce biases tied to the specific user base of each platform. For example, people with extremely positive or negative experiences are more inclined to post online reviews \cite{filieri2016makes}, which can impact the representation of the analysis. Future research could integrate data from a wider range of sources to reduce these biases.

Another avenue worth exploring is the inclusion of image-based data. For example, incorporating Google Maps Street View images and applying computer vision techniques \cite{kim2023examination} or multimodal language models could provide richer, more holistic insights into urban accessibility. This approach would allow for the automatic detection of physical barriers, such as stairs or narrow doorways, and the identification of accessible features like ramps or tactile paving. This could also enable us to make more meaningful recommendations for creating inclusive urban environments.

\section{Conclusions}

This study demonstrates the potential of leveraging crowdsourced data and advanced LLMs to assess public perceptions of urban accessibility. By analyzing Google Maps reviews across the United States using a fine-tuned Llama 3 model, we provide valuable insights into accessibility challenges at both the POI and socio-spatial levels. Our findings reveal that most POI types show negative sentiments, except for Recreation, with negativity mainly focused on service-related aspects. Our socio-spatial analysis shows that areas with higher proportions of white residents and greater socioeconomic status report more positive sentiment, while areas with more elderly and highly-educated residents express more negative views. However, no clear link is found between the presence of people with disabilities and public perception of accessibility. Overall, this approach offers a more comprehensive and nuanced understanding of accessibility issues compared to traditional survey methods, providing urban planners with crucial information to create more inclusive and accessible urban environments.

\bibliographystyle{ACM-Reference-Format}
\bibliography{sample-base}










\end{document}